\documentstyle[prl,aps,epsfig,multicol]{revtex}
\begin{document}
\def\be{\begin{equation}}
\def\ee{\end{equation}}
\def\bearr{\begin{eqnarray}}
\def\eearr{\end{eqnarray}}
\def\tc{$T_c~$}
\def\tcl{$T_c^{1*}~$}
\def\c2{ CuO$_2~$}
\def\ruo{ RuO$_2~$}
\def\lsco{LSCO~}
\def\bi{bI-2201~}
\def\tl{Tl-2201~}
\def\hg{Hg-1201~}
\def\sro{$Sr_2 Ru O_4$~}
\def\rc{$RuSr_2Gd Cu_2 O_8$~}
\def\mgb{$MgB_2$~}
\def\pz{$p_z$~}
\def\ppi{$p\pi$~}
\def\sqo{$S(q,\omega)$~}
\def\tperp{$t_{\perp}$~}

\title{RVB Contribution to Superconductivity in $MgB_2$}

\author{ G. Baskaran \\
Institute of Mathematical Sciences\\
C.I.T. Campus,
Madras 600 113, India }

\maketitle
\begin{abstract}

We view \mgb as electronically equivalent to (non-staggered) graphite 
($B^-$ layer) 
that has undergone a zero gap semiconductor to a superconductor 
phase transition by a large c-axis (chemical) pressure 
due to $Mg^{++}$ layers. Further, like the \ppi bonded planar organic 
molecules, graphite is an old resonating valence bond (RVB) system.  
The RVB's are the `preexisting cooper pairs' in the `parental' zero gap 
semiconducting $B^-$ (graphite) sheets that manifests themselves as 
a superconducting ground state of the transformed metal. Some 
consequences are pointed out.
\end{abstract}

\begin{multicols}{2}[]

Nature has once again surprised us through a
high temperature superconductivity in \mgb, discovered by
Akimitsu and collaborators\cite{aki1}.  This has resulted in a 
spurt of activity both in the 
experimental\cite{isotope,nmr,spht,arpes,neutron,aluminium,panaga,hall}
and theory\cite{bandstructure,correlation,furu} front. 
The aim of the present letter is to search for 
possible electronic contribution to the mechanism of
superconductivity in \mgb. Our theory uses the well recognized connection  
of electronic structure of boron sheets to graphite sheets\cite{aki1,bandstructure}.
In particular we view \mgb as a (non-staggered) graphite that has 
undergone a semi-metal 
to superconductor transformation because of a c-axis chemical pressure.  

The following are some of the phenomenological reasons for looking
for an electronic contribution to superconductivity in \mgb:
i) similarity to \ppi bonded planar systems like graphite, 
organic molecules and carbon nano-tubes where correlations play 
varying roles ii) absence of Hebel- Slichter peak in NMR 
relaxation\cite{nmr} iii) a temperature dependent peak around 17 meV 
in energy resolved neutron scattering
\cite{neutron} 
iv) first order metal to metal transition on Al or C substitution\cite{aluminium}
 v) possible anomalous temperature dependence of $R_H$\cite{hall} 
and also London penetration depth\cite{panaga} and vi) need to use an anomalously 
small  $\mu^*$ in phonon based theories\cite{mustar}
to get a reasonable \tc.

\mgb is very similar to graphite both crystallographically
and electronically. While the carbon hexagonal lattices are staggered
in graphite, the boron hexagonal layers  are on top of each 
other in \mgb.  The boron layers alternate with a triangular 
lattice of Mg layers. There is nearly 
complete 2e charge transfer from Mg to the boron sub system: 
\mgb $\equiv Mg^{++}(B^-)_2$.  Each boron acquires one electron and
acquires the electron configuration of a carbon atom:
\bearr
B^{-} (2s^2 2p^2) \equiv C (2s^2 2p^2) \nonumber
\eearr
Thus the $B^-$ sheets are electronically like graphite sheets.
The $Mg^{++}$ ion with its strong 
positive charge pulls the charged boron sheets closer and reduces 
the c-axis separation. The ratio of B-B distance along the ab plane 
to that along the c-axis is $\approx 2.0$ as opposed to the 
corresponding C-C distance ratio $ \approx 2.4 $ in graphite. 
{\em This 15 \% shortening of c-axis distance and removal of staggering 
converts a semi conducting `graphite' into a high Tc superconductor !} 

Graphite, like the \ppi bonded planar organic molecules, has been
the testing ground for the ideas of resonating valence bonds (RVB)
from early times. While Pauling\cite{pauling} and others emphasized the idea of
valence bond resonance and delocalization in the calculation of ground 
state (cohesive) energy and to some extent a quantitative understanding
of anomalous diamagnetism\cite{soos}, non trivial consequences of RVB ideas
for graphite have not been seriously discussed.  The aim of the 
present letter is to argue that RVB's, the preexisting enhanced singlet 
(cooper pair) correlations in graphite, a zero gap semiconductor
should reveal itself as a superconducting ground state after
an application of sufficient c-axis pressure (when no structural
modification intervenes). Nature seems to realize this through
a chemical pressure in \mgb, a system isoelectronic to graphite.  

In order to understand \mgb we should understand an isolated graphite
layer. The relevant valence orbital is the \pz orbitals of carbon 
with a mean occupancy of one electron. In the honeycomb structure 
there are two equivalent carbon atoms 
and two valence electrons per unit cell. According to band theory,
in view of two orbitals per unit cell, we get an empty 
and a filled band. Symmetry forces either a zero or a positive
overlap between the two bands. That is we either have a zero gap 
semiconductor or a semi metal. For a \ppi bonded system like graphite 
a well known model is the Hubbard model (simplified PPP model)
on a honeycomb lattice with nearest neighbor hopping: 
\be
H  =  -t \sum_{\langle ij \rangle } C^{\dagger}_{i\sigma} C^{}_{j\sigma} + h.c.
+ U\sum n_{i\uparrow}n_{i\downarrow}
\ee
The kinetic energy term gives us a zero gap semiconductor with a valence 
and conduction band dispersion 
\be
 \epsilon_{\pm}(k) = 
 \pm~ 2t 
 \left[ 
 \frac{3}{4} + \frac{1}{2} \cos k_x +
\cos \frac{k_x}{2} \cos \frac{ {\sqrt 3}k_y}{2}
\right]^{1\over2}
\ee
of the C layers with $t \approx 2.5 eV$.  The zero gap is not due to 
electron correlation; it is symmetry dictated. 
It disappears and produces small electron and hole fermi
pockets if we introduce next nearest neighbor hopping term in the planes 
or finite c-axis hopping term \tperp. Experimentally, in pure
graphite the electron and hole concentrations are rather small 
$\sim 10^{-4}$electron or hole per site; pure graphite 
sheets cleave and the binding between layers is van der Waals and not 
metallic.

It is known\cite{soos} that in \ppi bonded planar organic molecules the 
screened on site coulomb interaction in carbon 
(and also boron) \pz orbitals is $U \sim 6~eV$ (the bare atomic 
$U \sim 12~eV$), making the ratio ${U\over t} \sim 2.5$.  
According to early studies of Sorella and Tosatti\cite{sorella} and 
a very recent study of Furukawa\cite{furu} a Mott insulating behavior 
is obtained for a honeycomb lattice
only when $ {U\over t} \geq 3.5$.  Below this critical value of ${
U\over t}$ the zero gap character of the one electron states 
should make the many body effects at low energies less pronounced even 
though one is in 2 dimensions. However, some studies\cite{gr-tau,paco}, 
indicate anomalous life time for the quasi particles close to the 
fermi level in graphite.

The intermediate coupling character of the Hubbard model for graphite
makes it very difficult to approach it either as a weak coupling
or a strong coupling problem analytically. To circumvent this
problem, and also inspired by early RVB ideas on graphite, we propose 
an effective Hamiltonian for graphite/\mgb  on semi phenomenological 
grounds, and later sketch a possible microscopic derivation of this 
from the Hubbard model.

In the early treatment of graphite and \ppi bonded planar organic 
molecules, such as the discussion due to Pauling\cite{pauling}, the 
resonating valence bond (RVB) and enhanced valence bond amplitude 
is emphasized. Configurations with nearest neighbor singlet bonds
(VB) are encouraged in comparison to the polar (double
and single occupancy in the \pz orbitals) configurations. This concept
gave good estimate for cohesive energy, C-C bond distance and even some 
excited state properties, such as the singlet triplet exciton energy
differences. From the theoretical study\cite{pwascience,BZA,pwabook} 
of cuprates in the last one dozen years we have also realized that 
RVB's  are also preformed cooper pairs. They may have profound consequences 
such as high temperature superconductivity under appropriate conditions. 

Our primary aim now is to incorporate the well recognized RVB 
character of graphite in our effective Hamiltonian for \mgb.
With this in mind we propose the following model Hamiltonian 
for \mgb:
\bearr
H_{eff}  & \approx &
 -t  \sum_{\langle ij\rangle,\sigma} C^{\dagger}_{i,n\sigma} C^{}_{j,n\sigma} - 
  t_{\perp}\sum_{i,n,\sigma} C^{\dagger}_{i,n\sigma} C^{}_{i,{n+1}\sigma} + h.c.
\nonumber \\
&-& J \sum_{\langle ij\rangle, n} b^{\dagger}_{in,jn} b^{}_{in,jn} + 
- J_{\perp}\sum_{i,n} b^{\dagger}_{in,in+1} b^{}_{in,in+1} 
\eearr
Here i or j denotes a lattice point on the 2d honeycomb lattice and 
`n' is the layer index. The singlet operator 
$ b^{}_{ij} \equiv \frac{1}{\sqrt2} (C^{}_{i\uparrow}C^{}_{j\downarrow} 
- C^{}_{i\downarrow}C^{}_{j\uparrow})$. The two body term with 
$J_{ij} > 0 $  represents an energy gain when two electrons in 
neighboring sites 
i and j form a singlet (valence bond), as $b^{\dagger}_{ij} b^{}_{ij}$
is a number operator for bond singlets. That is, this term stabilizes 
covalent configurations relative to ionic configurations between any
two neighboring sites.  Tight binding fit of the states
close to the fermi level as obtained from ab-initio band structure 
calculation for \mgb give the values of the parameter  
$t \approx 1.6 eV, t_{\perp} \approx 1.25 eV$.

Now we outline an approximate microscopic derivation of the 
interaction term of our Model Hamiltonian (equation 3) and also 
estimate the phenomenological
parameters $J$ and $J_{\perp}$. In spirit it is similar to 
superexchange theory, but it is also designed to handle the
weak and intermediate coupling regions approximately. 

In graphite  as well as \mgb we are exactly at half filling 
in the sense of having one electron per \pz orbital. It is here we 
expect maximum effects of two body interactions (Hubbard U)
 in the ground state. 
In a free fermi gas in tight binding system, the spin state of two 
electrons in neighboring sites can be a singlet or a triplet. 
The local singlet character in k-space, as enforced by Pauli 
principle, does not enforce any type of singlet correlations in 
real space. However, two body collisions arising from the Hubbard
U term, encourages singlet amplitude between two electrons
of neighboring sites. This is the origin of the kinetic or
superexchange term in a Mott insulator at and close to half
filling.  

As in superexchange perturbation theory, we concentrate on two sites 
involving two electrons, but solve the two site problem exactly.
The two electron ground state is a singlet with an energy 
$E_g = -\frac{1}{2} \left[U^2 + 16t^2\right]^{1\over2} +\frac{U}{2}$. 
The energy of the triplet state is at $ E_T =0$. Two excited singlet 
states are at positive higher energies. In our effective Hamiltonian
(equation 3) the energy gain of a nearest neighbor singlet 
(valence or covalent) bond is represented by the term 
\be
- J b^{\dagger}_{ij}b^{}_{ij},{\hbox {~~~~~where~}} 
J = \frac{1}{2} \left[U^2 + 16t^2\right]^{1\over2} -
\frac{U}{2}
\ee
In the large $ t >> U$ limit the singlet stabilization 
energy $J \approx \frac{4t^2}{U}$, agreeing with the superexchange
perturbation theory.  

For \mgb using equation (4) we estimate $ J \approx 1.3 eV$
and $J_{\perp} \approx 0.4 eV$. Ideally, in the conducting metallic
state it is more meaningful to talk about residual coupling among 
excitations in momentum space. From this point of view the above
estimates of singlet stabilization terms in real space should be 
treated with caution. 

The valence bond stabilizing term is an attraction in the two body 
BCS singlet channel, near the fermi surface for our half filled case: 
\be
- J \sum_{\langle ij \rangle} b^{\dagger}_{ij}b^{}_{ij}
\rightarrow -J \sum 
C^{\dagger}_{k\uparrow}C^{\dagger}_{-k\downarrow}
C^{}_{-k'\downarrow} C^{}_{k'\uparrow}
\ee
This attraction, under normal conditions, will lead to a 
superconducting ground state. However, graphite is unusual in
the sense the density of states vanishes at the fermi level.
A simple cooper pair analysis shows us that the above attraction
is incapable of converting the zero gap semiconducting state into 
a superconducting state unless the attraction J exceeds a critical 
value $J_c \approx 3t$. Thus graphite does not have a superconducting
ground state, in spite of its RVB character ! 

\begin{figure}[h]
\epsfxsize 6cm
\centerline {\epsfbox{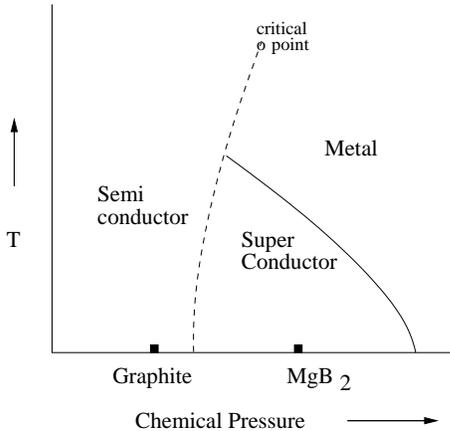}}
\caption{ Temperature - (c-axis) chemical pressure phase
diagram. Dashed line is the first order phase transition line
that ends in a critical point}
\end{figure}

Now we qualitatively discuss the zero gap semiconductor to 
superconductor transition when we go from graphite to \mgb.
As mentioned earlier, the ratio of the 
B-B distance along the c-axis to that in the plane gets reduced
from 2.4 to 2.0, when we go from graphite to \mgb. The coupling
along the c-axis is van der Waals like in graphite.  Early 
band structure calculations and transport studies have suggested  
a c-axis hopping matrix element \tperp $ \leq 0.25 eV$ for graphite.
The band structure results for \mgb show a large value
of \tperp $\sim 1.2 eV$ resulting in two large \ppi fermi surfaces 
and an addition of a small hole density to the
the 2 dimensional $\sigma$ bands of the honeycomb lattice.

Thus the major difference between graphite and \mgb is the c-axis
metallization resulting from the chemical pressure. This has given
rise to a finite density of states at the chemical potential 
for \mgb. Further, the local electronic structure and lattice
structure has not undergone any qualitative modification between 
graphite and \mgb. Thus \mgb, in view of its finite density of 
states at the fermi level is capable of utilizing the local RVB 
correlations and go into a superconducting state. 

The vanishing density of states due to the zero gap character 
in \mgb makes the valence bond stabilization term `irrelevant', 
as long as $ J < J_c \approx 3t $. That is, as a function of J there 
is a quantum phase transition from a zero gap semiconductor
to a superconductor. The value we have estimated for graphite 
$ J << 3t$,  is consistent with the experimental observation that 
graphite is not a superconductor.

A simple BCS mean field theory of our model Hamiltonian (equation 3)
gives the approximate formula for \tc
\be
k_BT_c \approx W e^{-{1\over J{\rho_{_0} }}},
\ee
Where W is the bandwidth of the \ppi band and $\rho_{_0}$ is the 
density of state at the fermi level. From the band structure 
calculations, there are two \ppi bands of width 18 and 13 eV.
We can take the mean bandwidth to be $ W \approx 15 eV$. 
The mean density of states arising from the two bands is 
$\rho_0 \approx {2\over 15}$ states per eV. If we assume a value of 
$J \approx 0.9 eV$ we get a \tc $\approx 40 K$. This suggests
that our renormalized $J \approx 1.3$, as obtained from our 
microscopic derivation is in the right ball park.

In our model, as we increase \tperp continuously from zero, 
the \tc increases continuously; in reality, with the increase of
c-axis pressure we expect a first order phase transition.
The zero gap semiconducting state has a screening which is 
fundamentally different from the metallic screening of a semi metal 
with a finite band overlap. On general grounds, using 
arguments similar to Mott, the metalization along the c-axis
arising from an uniaxial pressure will be a first
order phase transition. The c-axis lattice parameter collapse
arising from the c-axis metalization will add to this and make 
it a stronger first order transition like some of the known metal 
insulator transition.

We have sketched our general view on the zero gap semiconductor 
to superconductor transition schematically in figure 1 and 2.
In figure 1, there is a first order zero gap (or very small gap)
semiconductor to superconductor phase transition 
as a function of pressure in the ground state.

\begin{figure}[h]
\epsfxsize 9cm
\centerline {\epsfbox{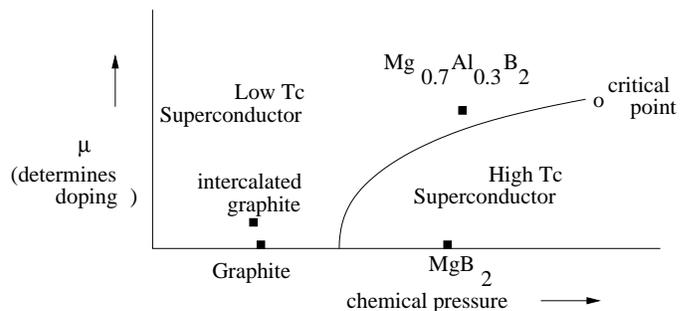}}
\caption{ Chemical potential ($\mu$) - (c-axis) chemical pressure
phase diagram. The line separating graphite and \mgb is a first order
phase boundary and it ends at a critical point}
\end{figure}

With a small addition of carriers (through Al substitution at
Mg site for example) we expect the first order phase transition
to survive as a function of chemical pressure. In figure 2, we have 
extended the phase diagram to the chemical potential (related to doping) 
vs chemical pressure (related to the c-axis lattice parameter) 
phase diagram. The first order line ends in a critical point. 

Experiments\cite{aluminium} on $Mg_{1-x}Al_xB_2$ and $MgB_{2-x}C_x$
have been performed for a range of x. In the case of Al substitution, 
\tc decreases gradually until x $\approx 0.1$  and for $0.1 < x <
0.25$ we have a two phase region. Superconductivity disappears
across the two phase region and the c-axis lattice parameter
jumps down by about 10 \%.  Al being trivalent adds an additional 
electron to the Boron subsystem. We believe that with Al or C 
substitution one crosses the first order phase boundary.

It is also known that intercalated graphites show low temperature
superconductivity. According to our picture intercalated graphites
never get a large fermi surface (and hence low \tc); 
as the average c-axis separation increases on intercalation, the
c-axis metalization does not occur. 

There are several issues that need to be understood both
qualitatively and quantitatively from our stand point. 
Further our theory has certain consequences which can be tested 
experimentally.

Symmetry of the order parameter is an important issue. Even in
cuprates, from the numerical analysis it is clear that
the d-wave and extended-s wave state are nearly degenerate.
In the present situation, where double occupancy is
not be completely projected in the low energy subspace, 
both s-wave and d-wave solutions are possible. The absence of
Hebel-Slichter peak in the NMR does indicate some similarity
to the cuprates and organic superconductors. The penetration
depth study Panagopoulos and collaborators\cite{panaga} also points 
to some gapless region. Further studies are necessary to
settle this issue.

There has been an intriguing neutron scattering result\cite{neutron}, 
where one sees two unexpected peaks at 17 meV and 30 meV in the 
phonon density of states. The 17 meV peak
in particular has a temperature dependence that peaks around
the \tc. As the authors have mentioned, is there a connection 
of this with the 41 meV peak of YBCO materials ? We suggest
that this peak is magnetic in origin and expresses an underlying
RVB character, and polarized neutron scattering experiment can
prove its spin 1 character. It is also interesting that this peak
starts building up around 150 K, indicating some kind of 
preformed pairing activity. It is also possible that the intrinsic \tc 
of \mgb is larger
$\approx 200 K$ and that the small carrier concentration in
the 2d $\sigma$ bond is acting like a source of dissipation
for superconductivity in the \ppi system, through interband 
scattering thereby reducing \tc considerably.
For us the lightly doped 2d $\sigma$ band, that is predicted
by band theory and used by electron and phonon based theories, 
does not play a crucial role in establishing high temperature 
superconductivity. As mentioned earlier, it might interferes 
with superconductivity in the \ppi band.

It is also interesting that the Hall resistivity $R_H$ has
a temperature dependence\cite{hall} below about 150 K. The strong
temperature dependence of $R_H$ in cuprates are known to
arise from the built up of RVB singlet correlations, 
suggesting a built up of RVB correlations in \mgb.

I thank R. Shankar (Madras) for discussions.

\end{multicols}
\end{document}